\documentclass[fleqn,usenatbib]{mnras}
\pdfoutput=1

\usepackage[T1]{fontenc}
\usepackage{ae,aecompl}
\usepackage{times} 
\usepackage{float} 
\usepackage{textcomp} 
\usepackage{natbib} 

\usepackage{graphicx}	
\usepackage{amsmath}	
\usepackage{amssymb}	

\title[Satellites of M33]{$\Lambda$CDM Predictions for the Satellite Population of M33}

\author[Patel et al.]{
Ekta Patel,$^{1}$\thanks{E-mail: ektapatel@email.arizona.edu} Jeffrey L. Carlin,$^{2}$ Erik J. Tollerud,$^{3}$ Michelle L. M. Collins$^{4}$  \newauthor \,\,and Gregory A. Dooley$^{5}$
\\
$^{1}$Department of Astronomy, The University of Arizona, 933 North Cherry Avenue, Tucson, AZ 85721, USA\\
$^{2}$Large Synoptic Survey Telescope,  950 North Cherry Avenue, Tucson, AZ 85719, USA\\
$^{3}$Space Telescope Science Institute, 3700 San Martin Drive, Baltimore, MD 21218, USA\\
$^{4}$Department of Physics, University of Surrey, Guildford, GU2 7XH, UK\\
$^{5}$Google, 111 8th Ave, New York, NY 10011, USA}

\date{Accepted XXX. Received YYY; in original form ZZZ}

\pubyear{2018}

\begin{document}
\label{firstpage}
\pagerange{\pageref{firstpage}--\pageref{lastpage}}
\maketitle

\begin{abstract}
Triangulum (M33) is the most massive satellite galaxy of Andromeda (M31), with a stellar mass of about $3\times10^9\; M_{\sun}$. Based on abundance matching techniques, M33's total mass at infall is estimated to be of order $10^{11}\; M_{\sun}$. $\Lambda$CDM simulations predict that M33-mass halos host several of their own low mass satellite companions, yet only one candidate M33 satellite has been detected in deep photometric surveys to date. This `satellites of satellites' hierarchy has recently been explored in the context of the dwarf galaxies discovered around the Milky Way's Magellanic Clouds in the Dark Energy Survey. Here we discuss the number of satellite galaxies predicted to reside within the virial radius ($\sim$160 kpc) of M33 based on $\Lambda$CDM simulations. We also calculate the expected number of satellite detections in N fields of data using various ground--based imagers. Finally, we discuss how these satellite population predictions change as a function of M33's past orbital history. If M33 is on its first infall into M31's halo, its proposed satellites are expected to remain bound to M33 today. However, if M33 experienced a recent tidal interaction with M31, the number of surviving satellites depends strongly on the distance achieved at pericenter due to the effects of tidal stripping. We conclude that a survey extending to $\sim$100 kpc around M33 would be sufficient to constrain its orbital history and a majority of its satellite population. In the era of WFIRST, surveys of this size will be considered small observing programs.

\end{abstract}



\section{Introduction}
Until the last decade, satellite galaxies had only been discovered around host galaxies approximately the mass of the Milky Way (MW) and Andromeda (M31) or greater. More advanced instruments and improved techniques for extracting stellar over-densities have now extended our view of satellites around host galaxies down to dwarf galaxy masses. For example, tens of dwarf satellite galaxies have recently been discovered around the Magellanic Clouds, the MW's most massive satellites \citep{dwagner15, koposov15, martin15, bechtol15, kim15a, kim15b}. This provides evidence that dwarf galaxy halos also harbor satellite galaxies of their own, supporting predictions from $\Lambda$ Cold Dark Matter ($\Lambda$CDM) numerical simulations \citep[e.g.,][]{moore99, gao04, kravtsov04, guo11, wang12, sales11, sales13}.

These discoveries have begun to fill in the faint end of the galaxy luminosity function, which is key to probing $\Lambda$CDM at small scales. $\Lambda$CDM is known for several challenges that arise upon comparing observations and $\Lambda$CDM simulations in the dwarf galaxy mass regime \citep[i.e. core-cusp problem, missing satellites, too big to fail; see][and references therein]{bullockbk17}. Adding to the observed sample of low stellar mass galaxies within the Local Group (LG) and even the Local Volume (< 10 Mpc) will therefore improve our understanding of $\Lambda$CDM and help determine whether these challenges are truly setbacks to the standard cosmological paradigm.

As the third most massive member of the LG, M33's stellar mass is roughly the same as the Large Magellanic Cloud (LMC) \citep{guo10, corbelli03}. Its gravitational influence on both its host galaxy and less massive galaxies in the LG is therefore non-negligible. The LMC and M33 are estimated to have dark matter halo masses of order $10^{11} \, M_{\sun}$ at infall based on various abundance matching (hereafter AM) techniques \citep[e.g.,][]{guo11}. However, M33 likely resides in a more massive dark matter halo than the LMC as evidenced by a peak circular velocity of 130 km s$^{-1}$ at 15 kpc \citep{corbellisalucci} compared to the LMC's peak circular velocity of about 92 km s$^{-1}$ \citep{vdmnk14}. If the LMC hosts several satellite galaxies, M33 is also expected to host its own satellite galaxies and perhaps even more due to a larger halo mass. Since M33 does not have a 1:10 stellar mass ratio binary companion like the Small Magellanic Cloud (SMC) is to the LMC, it is easier to study the halo of M33 and thereby its satellite population in detail. 

M33 can be considered an isolated analog of the LMC that resides at a close enough distance where even ultra-faint dwarf satellites \citep[$M_* < 10^5 \, M_{\sun}$ or $M_V > -7$; see][]{bullockbk17} can be detected. Observations of faint satellite galaxies around an M33 mass galaxy are crucial not only because of their importance to $\Lambda$CDM theory, but also because they may help us understand whether galaxies of this mass have a stellar halo. Stellar halos of MW mass galaxies are thought to be formed through the accretion of many satellite galaxies, but it is not yet clear whether this is true for less massive galaxies like the LMC and M33. Deep, wide-field observations at the distance of M33 may therefore provide insight on what the assembly histories for $\rm M_{DM} \sim 10^{11} \, M_{\sun}$ galaxies are (i.e. whether they are dominated by accretion events or in situ star formation).

Several ongoing surveys are searching for ultra-faint dwarf galaxy companions around the Magellanic Clouds. For example, the Magellanic Satellites Survey (MagLiteS) is using the Dark Energy Camera data to identify potential galactic companions of the Magellanic Clouds \citep{dwagner16, pieres17, torrealba18}. The Survey of the MAgellanic Stellar History \citep[SMASH;][]{nidever17} aims to find low surface brightness features around the Magellanic Clouds and has already identified a globular cluster that is likely associated with the LMC. Beyond the Local Group, the Magellanic Analog Dwarf Companions And Stellar Halos \citep[MADCASH;][]{carlin16} survey is observing isolated galaxies with masses similar to the Magellanic Clouds to map their halos and search for any associated dwarf companions. An extended survey around M33 would provide an additional counterpart to these growing samples. 

In this work, we predict the satellite population of M33 using the $\Lambda$CDM-based methodology of \citet[][hereafter D17A and D17B]{dooley17a,dooley17b} and provide observing strategies for two ground-based imagers to motivate a second search for the satellites of satellites phenomena in the LG. The main goal for predicting the M33 satellite population is to further constrain its orbital history and to explore alternative origins for its gaseous and stellar disk warps \citep{putman09,mcconnachie09} in the event that they cannot be explained by its past orbital history, as suggested in \citet{patel17a}. Since M33's past orbital history will directly affect the number of surviving satellites today, an extended survey around M33 can provide additional insight on M33's history. Finally, the orbital history of M33 is important to understanding the assembly of M31's halo, which appears to have had a very active recent accretion history \citep[e.g.,][]{fardal06, mackey10, ferguson16}. 

This paper is outlined as follows. In Section \ref{sec:observeddata}, we briefly discuss the mass of M33 and existing, deep observations of the M33 and M31 region. In Section \ref{sec:results}, we describe the methods of D17A and D17B, which make predictions for the satellite populations of host galaxies ranging in mass from the SMC to the MW. We then apply these methods to M33 to tabulate the number of satellites expected to reside within its virial radius. We also determine the number of those satellites that can potentially be observed for various survey radii given two different wide field imagers. Section \ref{sec:orbitalhistory} discusses different M33 orbital histories and how each would affect the surviving population of satellite companions. In Section \ref{sec:discussion}, we discuss the current morphology of M33, its potential origins in the context of satellite galaxy companions, and how M33 can improve our current understanding of $\Lambda$CDM at small scales. Finally, Section \ref{sec:conclusions} presents our conclusions.

\section{Observed Data}
\label{sec:observeddata}

\subsection{The Mass of M33}
As the third most massive member of the Local Group (LG), M33's mass and gravitational influence are important to understanding the LG's history. While M33's baryonic mass is fairly well constrained via observations, the total mass is less well known since the dark matter halo cannot be measured directly. Halo mass is the determining factor for how many satellite companions M33 can harbor, so we briefly discuss M33 mass estimates from the literature below. See Section 2 of \citet[][hereafter P17A]{patel17a} for more detailed explanations. 

\textbf{HI mass:} M33's extended gas disk was most recently observed by \citet[][hereafter P09]{putman09}. It extends to 22 kpc from the galaxy's center and its total HI content has a mass of about $1.4 \times 10^9 \, M_{\sun}$, approximately twice that of the LMC's HI disk. Most of the gas in the disk is located at radii beyond the extent of the star-forming disk. While it does not show any signs of significant truncation due to ram pressure, it does exhibit warps in both the north and south. These warps and their possible origin will be discussed further in Section \ref{sec:orbitalhistory}. 

M33's rotation curve was measured with 21-cm observations, illustrating that the circular velocity steadily rises out to 15 kpc \citep{corbellisalucci} where it is 130 km s$^{-1}$. This gives a dynamical mass ($V^2 = GM/r$) $\approx 5 \times 10^{10} \, M_{\sun}$. The continuous rise in the rotation curve may be linked to M33's warps at large radii. 

\textbf{Stellar mass:} M33's stellar content has been estimated by \citet{corbelli03} using rotation curve data and mass-to-light ratio arguments. This leads to a stellar mass range of $2.8-5.6 \times 10^9 \, M_{\sun}$. \citet{guo10} used M33's $B-V$ color and stellar population models to estimate M33's mass-to-light ratio giving a stellar mass of $2.84 \pm 0.73 \times 10^9 \, M_{\sun}$. Combining these estimates into one result gives $3.2 \pm 0.4 \times 10^9 \, M_{\sun}$ \citep{vdm12ii}. We adopt this value for the remainder of this analysis, though our results are most sensitive to M33's adopted halo mass.

\textbf{Halo mass:} M33's baryonic mass content sums to approximately $6.4\times 10^9 \, M_{\sun}$ \citep{corbelli03}. Dividing this by the dynamical mass gives a baryon fraction of 12.8\%. It has been shown that only a fraction of cosmic baryons condense to form stars in spiral galaxies, resulting in low baryon fractions between 3-5\% \citep[e.g.,][]{trujillogomez15, rodriguezpuebla12, trujillogomez11, leauthaud11, behroozi10, guo10, jiang07, mandelbaum06, fukugita98}. Applying this baryon fraction to M33 suggests its mass at infall could have been $1.3-2.1 \times 10^{11} \, M_{\sun}$.  At this halo mass scale, M33 is approximately 10\% of M31's mass. More importantly, $\Lambda$CDM predicts that all $10^{11} \, M_{\sun}$ halos host a few satellites with $M_* > 10^3 \, M_{\sun}$ across independent AM techniques (see D17A). Details of M33's predicted satellite population will be discussed in Section \ref{sec:results}. 

\subsection{Optical Observations of the M33-M31 Region}
Numerous multi-wavelength observations have been taken  of the M33-M31 region. Below we summarize one particular imaging survey of resolved stars in the M31 system that has led to many discoveries of new dwarf galaxies, globular clusters, and stellar streams in M31's extended halo. 

\begin{figure}
    \includegraphics[width=1.1\columnwidth, trim=12mm 12mm 0mm 12mm]{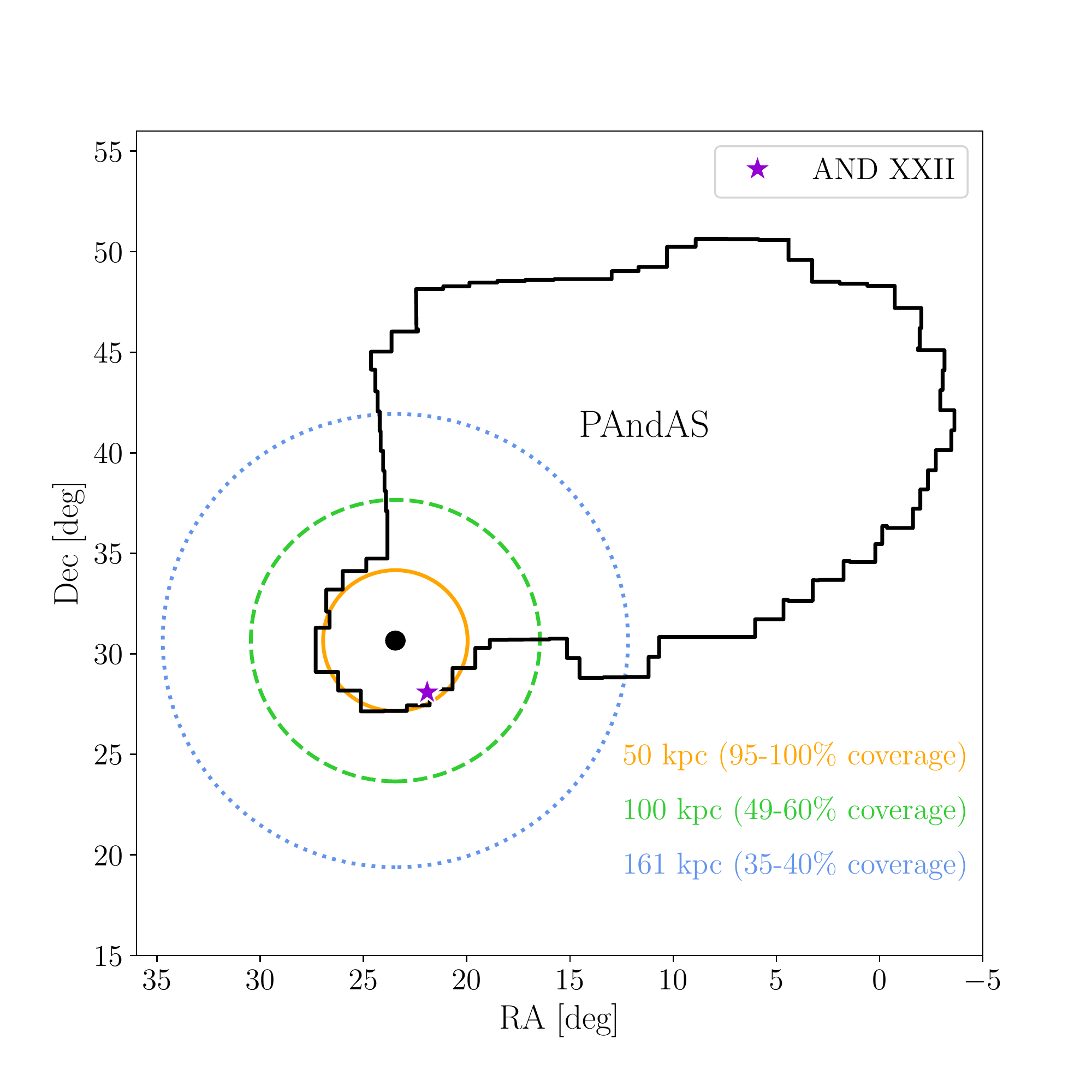}
    \caption{The PAndAS footprint (black) compared to the virial radius of M33 (dotted blue circle), a 100 kpc region around M33 (dashed green circle), and a 50 kpc region around M33 (orange). All circles assume as distance of 794 kpc to M33. The closed black circle indicates three scale lengths of M33's stellar disk using a scale length of approximately 0.15$^{\circ}$. The fraction of the virial volume associated with each circle that is encompassed by the PAndAS footprint is listed on the bottom right. The lower limit comes from a distance of 794 kpc to M33 and the upper limit assumes a distance of 968 kpc to M33. The position of And XXII, the only potential M33 satellite known to date is indicated with a purple star.}
    \label{fig:coverage}
\end{figure}

The Pan-Andromeda Archaeological Survey (PAndAS) used the 1 square degree field of view of MegaPrime/MegaCam on the 3.6m Canada-France-Hawaii Telescope \citep[CFHT,][hereafter M09]{mcconnachie09}. The survey footprint is illustrated in black in Fig. \ref{fig:coverage}. The survey covers more than 300 square degrees (equivalent to 70,000 kpc$^2$) spanning the region extending to 150 projected kpc ($10.5^{\circ}$) from the center of M31 and 50 projected kpc ($3.5^{\circ}$) from the center of M33. Stars in M31 are resolved down to $g=26.5$ and $i=25.5$. 

These data have led to the discovery of new M31 stellar structures in the northwest and southwest, which both extend to 100 kpc, and the extension of a previously known stream in the east. Sixteen additional dwarf galaxies \citep[see][and references therein, hereafter M16]{martin16} and 59 globular clusters \citep{huxor14} were also identified in the PAndAS data. PAndAS and preceding surveys have greatly increased our view of substructures in M31's inner and outer halo. While the faintest and most diffuse structures within this survey may fall below the completeness limits, PAndAS has revolutionized our view of the M31 system thus far. 

Of the dwarf galaxies discovered in the PAndAS survey, only one is considered a candidate satellite of M33 rather than M31. And XXII lies to the southwest of M33 on the sky and its relative systemic velocity, angular separation from M33, and distance all indicate that it may be a satellite of M33 instead of M31 \citep{tollerud12, chapman13, martin09}. The position of And XXII is denoted by a purple star in Fig. \ref{fig:coverage}. 

\citet{chapman13} present the results of an N-body simulation of M33 and M31 from which they conclude that during a close pericentric passage, only M33 halo particles at large distances from M33 would be stripped. Assuming that bound satellites follow the dark matter distribution of their host, And XXII would remain bound to M33 during such an encounter based on these simulations. These results strongly depend on the adopted M33 orbital history and other potential orbits may not yield the same conclusion for And XXII.

The absolute magnitude of And XXII  is $M_V =-6.5$\footnote{M16 report $M_V =-6.7$ for And XXII, resulting in $M_* \approx 4.6\times10^4 \, M_{\sun}$. Our conclusions are unaffected by this incremental difference in And XXII's approximate stellar mass.} and M33's $M_V=-18.8$ \citep{mcconnachie12}. Using these values, a rough approximation for the stellar mass content of And XXII can be derived. If we adopt an M33 stellar mass of $3.2\times10^9 \, M_{\sun}$ and assume that the stellar mass-to-light ratio is unity for ultra-faint dwarfs as in \citet{mcconnachie12}\footnote{A ratio of unity allows one to easily scale to any preferred stellar mass-to-light ratio.}, then And XXII has a stellar mass $M_* \approx 3.8\times10^4 \, M_{\sun}$. This means the PAndAS survey is sensitive to galaxies near the upper end of the ultra-faint dwarf regime \citep[see][]{bullockbk17}. This stellar mass matches to a virial mass of $\gtrsim 5\times10^8 \, M_{\sun}$ according to the \cite{moster13} AM technique. 

 Satellites of M31 have been discovered in the Sloan Digital Sky Survey (SDSS) \citep{zucker04, zucker07}, which reaches a limiting magnitude of $M_V=-8$ at best for a distance of 800-1000 kpc \citep{tollerud08}. The SDSS photometry is roughly sensitive to galaxies with stellar mass $M_* \gtrsim 1.5\times 10^5 \, M_{\sun}$ assuming a mass-to-light ratio of unity. The stellar mass limits quoted in this work should be considered lower limits as a function of survey depth because the stellar mass-to-light ratios of the ultra-faint dwarfs are more consistent with ratios between one and two \citep{kirby13, mcgaugh13, mcgaugh10, tollerud11a, woo08}. For example, \citet{kirby13} assumes $M_*/L_V=1.6 M_{\sun}/L_{\sun}$ based on the average values for dwarf spheroidals in \citet{woo08}.

\section{Methods and Results} 
\label{sec:results}

In this section, we summarize the methods of D17A and D17B to derive the predicted satellite population of an M33-mass galaxy. Given that M33's virial radius is significantly larger than the area surveyed by PAndAS, we motivate the need to complete the search for satellites around M33 using available and upcoming wide field imagers. The average number of observed satellites that M33 is expected to host is reported for two different instruments and for a range of M33 distances following the analysis of D17A and D17B. 

\subsection{D17 Predictions for Luminous Satellites}
\label{subsec:luminoussats}
In D17A , the \textit{Caterpillar} simulation suite \citep{griffen16} is used to predict the number of satellites expected to exist around MW mass halos and lower mass field halos in a $\Lambda$CDM paradigm. Since \textit{Caterpillar} is a suite of dark matter only simulations, several additional steps are taken to include the effects of reionization and to assign stellar masses to dark matter subhalos. Here we briefly summarize the the \textit{Caterpillar} simulation suite and the methods of D17A. 

\textit{Caterpillar} is a suite of 33 high particle resolution ($m_p = 3\times10^{4} \, M_{\odot}$) and high temporal resolution (5 Myr/snapshot) zoom-in simulations of MW mass galaxies. The suite was run with \ion{}{P-GADGET3} and \ion{}{GADGET4}, which are tree-based N-body codes derived from \ion{}{GADGET2} \citep{springel05}. \textit{Caterpillar} adopts the \textit{Planck} cosmology: $\Omega_m=0.32$, $\Omega_{\Lambda}=0.68$, $\Omega_b=0.05$, $n_s=0.96$, $\sigma_8=0.83$, and Hubble constant $H_0=67.11$ km s$^{-1}$ Mpc$^{-1}$ \citep{planck14}. Dark matter halos are identified using the \ion{}{ROCKSTAR} \citep{rockstar} algorithm and merger trees are created with \ion{}{CONSISTENT-TREES} \citep{behroozi13a}. \ion{}{ROCKSTAR} assigns virial masses to halos based on the \citet{brynorman98} definition.

Using the 33 simulated \textit{Caterpillar} halos, subhalo mass functions (SHMFs) are computed for several AM methods explored in D17A. The SHMFs follow the general form 

\begin{equation}
\rm \frac{dN}{dM_{sub}} = K_0 \left( \frac{M_{sub}}{M_{\sun}}  \right)^{-\alpha} \frac{M_{host}}{M_{\sun}}, 
\label{eq:SHMF}
\end{equation}

where $K_0$ and $\alpha$ depend on halo mass. For the \citet[][hereafter GK17]{gk17a} AM model, $\alpha=1.82$ and $K_0=0.000892$ for field halos and $\alpha=1.87$ and $K_0=0.000188$ for MW analogs. This SHMF accounts for all self-bound subhalos within $R_{\rm vir}$ at z=0, however subhalos of subhalos are excluded. Ideally, these sub-subhalos will be included in this type of analysis in the future. We will use the D17A and D17B results for the GK17 AM method in the remainder of this analysis. 

Given the SHMF, the mean number of dark matter subhalos predicted to exist within the virial radius of a host galaxy with mass $\rm M_{host}$ is calculated by integrating Eq. \ref{eq:SHMF} from $\rm M_{min}$ to $\rm M_{\rm host}$.  This yields:

\begin{equation}
\rm \bar{N} = \frac{K_0 M_{host}}{\alpha-1} \left(M_{min}^{1-\alpha} - M_{host}^{1-\alpha}\right). 
\label{eq:meanN}
\end{equation}

D17A chooses $\rm M_{\rm min}=10^{7.4} \, M_{\sun}$ as a threshold below which no star formation occurs due to reionization. Using Eq. \ref{eq:meanN}, they then generate random realizations of $\bar{N}$ according to a Poisson distribution and randomly assign halo masses based on the SHMF (Eq. \ref{eq:SHMF}). Once the halo masses are assigned, the reionization model is implemented.  

Reionization is accounted for by randomly assigning whether simulated halos should host stars or remain dark according to \citet{barber14}. This relation depends on the host halo's mass. The fraction of halos that are luminous at z=0 is shown in D17A, Figure 3. Approximately 50\% of halos with $\rm M_{vir,peak} > 10^9 \, M_{\sun}$ are luminous. This function is based on a semi-analytic model in which reionization occurs from $z=15$ to $z=11.5$ \citep{starkenburg13}. Since D17A considers several different AM methods, halos are determined to be dark or luminous halos with respect to the appropriate mass definition. In the GK17 AM method, stellar masses are assigned using the peak virial mass of halos. All mentions of virial quantities here also refer to the \cite{brynorman98} definition. In D17A, reionization is implemented as an instantaneous process at z=13.3.

Finally, halos are assigned stellar masses using the $M_* - M_{\rm halo}$ relation for the AM model of choice. For the GK17 AM model, the scatter in assigning stellar masses is explicitly taken into account. All simulated halos with galaxy stellar masses above $10^3 \, M_{\sun}$ are included.

The mean predicted number of luminous satellites ($\rm \bar{N}_{lum}$) around the LMC and the SMC are provided in Figure 1 of D17B as a function of satellite stellar mass. A stellar mass of $2.6 \times 10^9 \, M_{\sun}$ is adopted for the LMC, which translates to $M_{\rm vir} = 2.3 \times 10^{11} \, M_{\sun}$ using the GK17 AM method. This virial mass corresponds to a virial radius of $R_{\rm vir}=156$ kpc. The stellar mass of M33 is approximately $3.2\times10^9 \, M_{\sun}$, giving a virial mass  $M_{\rm vir} = 2.5 \times 10^{11} \, M_{\sun}$ and $R_{\rm vir}=161$ kpc.  Throughout this analysis, we adopt this virial mass and radius for M33, however, other AM methods \citep{moster13, behroozi13b, brook14, gk14} yield slightly different masses, virial radii, and therefore different satellite populations as discussed in D17A. Since $M_{\rm vir, M33}/M_{\rm vir, LMC} \approx 1.09$ and $\rm \frac{dN}{dM_{sub}} \propto M_{\rm host}$, one can take the results for the LMC in D17B and scale them up by $\sim$9\% to acquire the predicted number of luminous satellites for M33. These results are represented by the solid gray line and shaded region in Fig. \ref{fig:794kpc}.

We will refer to the combined methodology of D17A and D17B as \textit{D17 predictions} throughout the remainder of this work. This refers to the aggregate of $\Lambda$CDM predictions (from the \textit{Caterpillar} suite of simulations), the GK17 AM method, and the reionization model described above.

\subsection{Correcting for Geometric Effects}
\label{subsec:geometry}
Thus far, we have described how D17A and D17B used $\Lambda$CDM simulations to compute the mean number of luminous satellites in the virial volume of isolated galaxies with masses similar to M33 and the LMC. However, these numbers do not directly translate into expectations for observations which depend on the field of view, distance to the target galaxy, and the virial radius of the target galaxy. These predictions must therefore be corrected (using a multiplicative scaling factor) for the number of satellites expected within the line of sight for a given survey radius. This scaling factor from D17A will be referred to as $K_{\rm los}(R)$.

$K_{\rm los}(R)$ is the fraction of satellites expected to exist within a cylinder along the line of sight centered on the host galaxy that includes the radial dependence of the satellite galaxy distribution. Therefore multiplying $K_{\rm los}(R)$ by the number of satellite galaxies predicted to reside within $R_{\rm vir}$ of a galaxy by the D17 predictions (i.e. the gray line in Fig. \ref{fig:794kpc}) yields the number of observed satellites expected in a given field of view (i.e. the colored lines in Fig. \ref{fig:794kpc}). 

$K_{\rm los}(R)$ is derived as follows. Eq. \ref{eq:K} provides a fit to the radial distribution of satellites galaxies where satellite galaxies are the subhalos  from the {\em Caterpillar} simulations that are determined to be luminous via the steps outlined in Section \ref{subsec:luminoussats}.

\begin{equation}
K(R) =  \left\{
        \begin{array}{ll}
            k_1R + k_2R^2 + k_3R^3, \,\,\, R < 0.2 \\
            k_4\arctan\left(\frac{R}{k_5} - k_6\right), \, 0.2 \leq R \leq 1.5
        \end{array}
    \right.
\label{eq:K}
\end{equation}

where $k_1=-0.2615$, $k_2=6.888$, $k_3=-7.035$, $k_4=0.9667$, $k_5 =0.5298$, and $k_6=0.2055$. 

Physically, the volume we are interested in considering is a cylinder of radius $R$ and half-depth $Z$ inscribed within a sphere of radius $R_{\rm vir}$ given the D17 formalism. $R \equiv r/R_{\rm vir}$ such that $R=1$ corresponds to the virial radius of a galaxy's halo, which is approximated as a sphere. In practice, $r$ would be the field of view radius of a survey, $\rm R_{FoV}$. The function $\frac{1}{4\pi R^2} \frac{dK}{dR}$ then describes the density of satellite galaxies as a function of $R$ (or $r$ in spherical coordinates). Integrating this density function in spherical coordinates over a cylinder gives the scaling factor, $K_{\rm los}(R)$. The density function should be integrated using bounds corresponding to a cylinder with a radius of $R$ and a half-depth of $Z \equiv z/ R_{\rm vir}$. $K_{\rm los}(R)$ is plotted for values of $Z=1$ and $Z=1.5$ in Figure 11 of D17A. 

For our calculations, we follow the conventions of D17A and adopt $Z=1.5$, which corresponds to approximately the splashback radius of the target host galaxy, or the distance at which bound satellites will reach their first apocenter \citep{more15}. Beyond this radius, the density of satellites is expected to rapidly decrease to zero.

\subsection{The Predicted Satellite Population of M33}
\label{subsec:fullsurvey}

\begin{table*}
	\centering
	\caption{Expectations for the number of satellites that could be observed with Hyper Suprime-Cam (HSC)  in the area corresponding to $\rm N_{fields}$, or the total number of fields observed. The first row shows the D17 predictions for the number of satellites within $\rm R_{vir}$. Column 3 lists $\rm R_{FoV}$ or the radius of the projected area observed by $\rm N_{fields}$ at the distance of M33. Columns 4-7 show the expected number of observed satellites at different limiting stellar masses using the scaling factor $K_{\rm los}(R)$. HSC's field of view has a diameter of 1.5$^{\circ}$. The corresponding physical radius for a single field is approximately 10.4, 11.5, and 12.7 kpc at M33 distances of 794, 880, 968 kpc respectively. All calculations assume a 100\% observational efficiency rate.}
	\begin{tabular}{p{3cm}cccccc}  
		\hline \hline
	   &  & &  $\rm \bar{N}_{lum} \; (M_* > 10^3 M_{\sun})$ &$\rm \bar{N}_{lum} \; (M_* > 10^4 M_{\sun})$ &$\rm \bar{N}_{lum} \; (M_* > 10^5 M_{\sun})$ &$\rm \bar{N}_{lum} \; (M_* > 10^6 M_{\sun})$ \\ \hline 
	   
D17 predictions   &&  & $11.42^{+3.86}_{-3.82}$ & $8.07^{+3.16}_{-3.18}$ & $4.04^{+2.13}_{-2.20}$ & $1.45^{+1.27}_{-1.28}$ \\ 

	($R_{\rm survey} = R_{\rm vir}$) &&&&&&  \\ \hline \hline

   \textbf{ D$\rm_{M33}$ = 794 kpc }& $\rm N_{fields}$ & $\rm R_{FoV}$ (kpc) &  $\rm \bar{N}_{obs} \; (M_* > 10^3 M_{\sun})$ &  $\rm \bar{N}_{obs} \; (M_* > 10^4 M_{\sun})$ &$\rm \bar{N}_{obs} \; (M_* > 10^5 M_{\sun})$ &$\rm \bar{N}_{obs} \; (M_* > 10^6 M_{\sun})$ \\ \hline
   
								 & 1 &10.4  & 0.53 & 0.37 & 0.19 & 0.07 \\ \hline
  							      & 10 & 32.9 & 3.59 & 2.53 & 1.27 & 0.46 \\ \hline
							      & 23 & 49.8 & 5.86 & 4.14 & 2.07 & 0.74 \\ \hline
							      & 50 & 73.5 & 8.25 & 5.83 & 2.92 & 1.05 \\ \hline
							      & 100 & 103.9 & 10.32 & 7.29 & 3.65 & 1.31 \\ \hline
							      & 154 & 129.0 & 11.42 & 8.07 & 4.04 & 1.45 \\ \hline \hline       

   \textbf{ D$\rm_{M33}$ = 880 kpc } & $\rm N_{fields}$ &  $\rm R_{FoV}$ (kpc) & $\rm \bar{N}_{obs} \; (M_* > 10^3 M_{\sun})$ &  $\rm \bar{N}_{obs} \; (M_* > 10^4 M_{\sun})$ &$\rm \bar{N}_{obs} \; (M_* > 10^5 M_{\sun})$ &$\rm \bar{N}_{obs} \; (M_* > 10^6 M_{\sun})$ \\ \hline
                                   & 1 &  11.5 & 0.64 & 0.45 & 0.23 & 0.08  \\ \hline
                                   & 10 & 36.4 & 4.11 & 2.90 & 1.45 & 0.52 \\ \hline
                                   & 23 & 55.2 & 6.48 & 4.58 & 2.29 & 0.82 \\ \hline
                                   & 50 & 81.4 & 8.88 & 6.28 & 3.14 & 1.13 \\ \hline
                                   & 100 & 115.2 & 10.88 & 7.69 & 3.85 & 1.38 \\ \hline
                                   & 125 & 128.8 & 11.42 & 8.07 & 4.04 & 1.45 \\ \hline \hline 
  
  \textbf{ D$\rm_{M33}$ = 968 kpc }& $\rm N_{fields}$ & $\rm R_{FoV}$ (kpc) & $\rm \bar{N}_{obs} \; (M_* > 10^3 M_{\sun})$ &  $\rm \bar{N}_{obs} \; (M_* > 10^4 M_{\sun})$ &$\rm \bar{N}_{obs} \; (M_* > 10^5 M_{\sun})$ &$\rm \bar{N}_{obs} \; (M_* > 10^6 M_{\sun})$ \\ \hline
								 & 1 & 12.7 & 0.77 & 0.54 & 0.27 & 0.10 \\ \hline
								 & 10 & 40.1 & 4.61 & 3.26 & 1.63 & 0.59 \\ \hline
								 & 23 & 60.8 & 7.07 & 4.99 & 2.50 & 0.90 \\ \hline
								 & 50 & 89.6 & 9.46 & 6.68 & 3.35 & 1.20 \\ \hline
								 & 103 & 128.6 & 11.42 & 8.07 & 4.04 & 1.45 \\ \hline \hline
	\end{tabular}
	\label{table:HSC}
\end{table*}
\begin{table*}
	\centering
	\caption{Similar to Table \ref{table:HSC} for the MegaCam/MegaPrime (MC) imager. MC has a square field of view with a 1$^{\circ}$ diameter. We have approximated it as a circular field of view with the same area (or a $1.13^{\circ}$ diameter). $R_{\rm FoV}$ for a single field corresponds to 7.8, 8.7, and 9.5 kpc at M33 distances of 794, 880, 968 kpc respectively. 
	}
	\begin{tabular}{p{3cm}cccccc}  
		\hline \hline
	   &  &  &  $\rm \bar{N}_{lum} \; (M_* > 10^3 M_{\sun})$ &$\rm \bar{N}_{lum} \; (M_* > 10^4 M_{\sun})$ &$\rm \bar{N}_{lum} \; (M_* > 10^5 M_{\sun})$ &$\rm \bar{N}_{lum} \; (M_* > 10^6 M_{\sun})$ \\ \hline 
	   
D17 predictions   &&  & $11.42^{+3.86}_{-3.82}$ & $8.07^{+3.16}_{-3.18}$ & $4.04^{+2.13}_{-2.20}$ & $1.45^{+1.27}_{-1.28}$ \\ 

	($R_{\rm survey} = R_{\rm vir}$) && &&&&  \\ \hline \hline

   \textbf{ D$\rm_{M33}$ = 794 kpc }& $\rm N_{fields}$ &  $\rm R_{FoV}$ (kpc) &$\rm \bar{N}_{obs} \; (M_* > 10^3 M_{\sun})$ &  $\rm \bar{N}_{obs} \; (M_* > 10^4 M_{\sun})$ &$\rm \bar{N}_{obs} \; (M_* > 10^5 M_{\sun})$ &$\rm \bar{N}_{obs} \; (M_* > 10^6 M_{\sun})$ \\ \hline
   								 & 10 & 24.8 & 2.38 & 1.68 & 0.84 & 0.30 \\ \hline
								 & 41 & 50.1 & 5.90 & 4.17 & 2.09 & 0.75 \\ \hline
								 & 100 & 78.3 & 8.64 & 6.11 & 3.06 & 1.10 \\ \hline
								 & 200 & 110.7 & 10.67 & 7.54 & 3.77 & 1.35 \\ \hline
								 & 254 & 128.9 & 11.42 & 8.07 & 4.04 & 1.45 \\ \hline \hline

   \textbf{ D$\rm_{M33}$ = 880 kpc }&  $\rm N_{fields}$ &  $\rm R_{FoV}$ (kpc) &$\rm \bar{N}_{obs} \; (M_* > 10^3 M_{\sun})$ &  $\rm \bar{N}_{obs} \; (M_* > 10^4 M_{\sun})$ &$\rm \bar{N}_{obs} \; (M_* > 10^5 M_{\sun})$ &$\rm \bar{N}_{obs} \; (M_* > 10^6 M_{\sun})$ \\ \hline								 
	   						     & 10 & 27.4 & 2.77 & 1.96 & 0.98 & 0.35 \\ \hline
							     & 41 & 55.6 & 6.52 & 4.61 & 2.31 & 0.83 \\ \hline
							     & 100 & 86.8 & 9.27 & 6.55 & 3.28 & 1.18 \\ \hline
							     & 200 & 122.7 & 11.20 & 7.92 & 3.96 & 1.42 \\ \hline
							     & 207 & 128.7 & 11.42 & 8.07 & 4.04 & 1.45 \\ \hline \hline
							     
   \textbf{ D$\rm_{M33}$ = 968 kpc }& $\rm N_{fields}$ &  $\rm R_{FoV}$ (kpc)&  $\rm \bar{N}_{obs} \; (M_* > 10^3 M_{\sun})$ &  $\rm \bar{N}_{obs} \; (M_* > 10^4 M_{\sun})$ &$\rm \bar{N}_{obs} \; (M_* > 10^5 M_{\sun})$ &$\rm \bar{N}_{obs} \; (M_* > 10^6 M_{\sun})$ \\ \hline								 
	   						     & 10 & 30.2 & 3.18 & 2.25 & 1.13 & 0.40\\ \hline
							     & 41 & 61.1 & 7.10 & 5.02 & 2.51 & 0.90 \\ \hline
							     & 100 & 95.4 & 9.83 & 6.95 & 3.48 & 1.25 \\ \hline
							     & 171 & 128.8 & 11.42 & 8.07 & 4.04 & 1.45\\ \hline \hline
							   
	\end{tabular}
	\label{table:Mega}
\end{table*}

Fig. \ref{fig:coverage} shows the PAndAS survey footprint in black along with three circles indicating survey regions of different radii centered on M33. The orange circle represents an area with a radius of 50 kpc, approximately the same as the PAndAS coverage around M33 if the bridge between M33 and M31 is excluded. The dashed green circle indicates the area spanned by a radius of 100 kpc. The PAndAS survey already observed $\sim$50\% of this region. The dotted blue circle represents the area contained within the adopted virial radius of M33, or 161 kpc. Only $\sim$40\% of this region is contained within the PAndAS survey. A comparison of the orange and blue areas reveals the drastic difference in area coverage of available data relative to the true extents of M33's full halo, thereby motivating the need to search for satellites in the outskirts of M33. 

The PAndAS survey used MegaCam/MegaPrime (MC) on the CFHT, which has a $1^{\circ} \times 1^{\circ}$ square field of view. For our calculations, we have approximated it as a circular field of view with a $1.13^{\circ}$ diameter or a circular field with the same area. Hyper Suprime-Cam (HSC) on the 8.2m Subaru telescope has a field of view with a $1.5^{\circ}$ diameter. Other imagers, such as the Dark Energy Camera on the Blanco telescope and LSST would be ideal for surveys around M33; however, they are located in the South and will not observe the M31 region.

In Tables \ref{table:HSC} and \ref{table:Mega}, we tabulate how many luminous satellites are expected to exist around M33 for various satellite mass (or limiting magnitude) thresholds. Each table corresponds to a different instrument (HSC or MC). The first row in both tables lists the mean number of luminous satellites, $\rm \bar{N}_{lum}$, M33 is expected to host within its virial radius from the D17 predictions. We list the corrected number of satellites, $\bar{N}_{\rm obs}$ in the following rows as a function of the number of observed fields, $\rm N_{fields}$. The values in these rows are calculated by multiplying the predicted number of luminous satellites $\rm \bar{N}_{lum}$ by the scaling factor $K_{\rm los}(R)$, which is calculated using $\rm R = ( N_{fields} \times R_{FoV}^2/R_{vir}^2) ^{1/2}$ and Z=1.5. $\rm R_{FoV}$ is the physical radius (in kpc) that corresponds to the radius observed by $\rm N_{fields}$ at the distance of M33 for each imager and $\rm R_{vir}=161$ kpc. Several rows are included from $\rm N_{fields}=1$ to the maximum number of fields needed to detect all potential satellites using HSC or MC. 

Distance measurements to M33 in the literature range from 794 kpc \citep{mcconnachie04} to 968 kpc \citep{u09}; thus we repeat all calculations for M33 distances of 794 kpc, 880 kpc, and 968 kpc. For M33 distances > 794 kpc, fewer fields are required for both HSC and MC. Note that all results assume a $\Lambda$CDM cosmology and that the observational efficiency rate for any number of fields is 100\% (i.e. if 10 HSC fields are taken, then two satellites with $M_* > 10^3 \, M_{\sun}$ are guaranteed at a distance of 794 kpc). The number of fields, $\rm N_{fields}$, listed are approximate and do not account for the fact that off-axis fields probe a smaller volume than on-axis fields. They also do not account for the overlapping regions between fields.  

Since $K_{\rm los}(R)$ is approximately zero at $R\equiv R_{\rm fov}/R_{\rm vir} \lesssim 0.05$, Table \ref{table:Mega} excludes $\rm N_{fields}=1$. The smaller field of view of MC requires significantly more fields to probe the area around M33, thus, it is not surprising that the PAndAS survey did not extend to larger radii. The satellite stellar mass thresholds range from $10^3-10^6 \, M_{\sun}$, or from the ultra-faint dwarf regime to the classical dwarf satellite regime. According to D17A, the probability that an M33-stellar mass galaxy hosts a satellite with $M_* > 10^3, 10^4$, and $10^5 \, M_{\sun}$ is unanimously > 95\% for the GK17 AM method. 

\begin{figure*}
\includegraphics[scale=0.8]{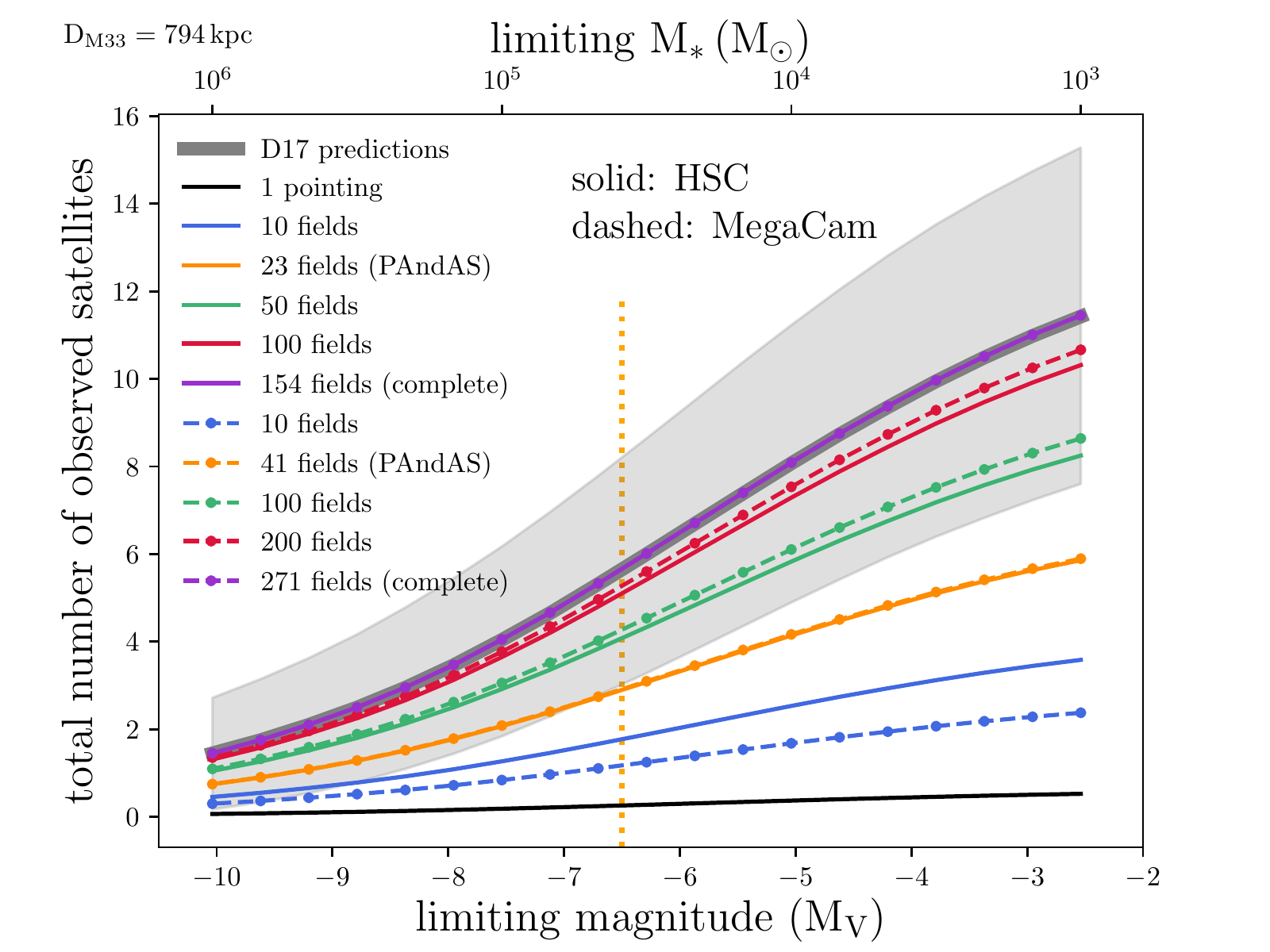}
\caption{The total number of predicted M33 satellites as a function of limiting visual magnitude and limiting stellar mass. A mass-to-light ratio of unity is used to convert from $\rm M_V$ to $\rm M_*$. The gray solid line and shaded region represent the D17 predictions. All colored lines refer to the D17 predictions corrected using the $K_{\rm los}$ scaling factor. Each line corresponds to N fields observed with HSC (solid lines) or MC (dot dashed lines). The PAndAS area is represented by 23 fields using HSC (solid orange line) and 41 fields using MC (dot dashed orange line). These two lines are coincident since they correspond to the same area. These results are exclusively for an M33 distance of 794 kpc. The dotted vertical orange line indicates the faintest object detected in the PAndAS survey.} 
\label{fig:794kpc}
\end{figure*}

Fig. \ref{fig:794kpc} summarizes the results in Tables \ref{table:HSC} and \ref{table:Mega}, illustrating the number of expected satellites not only as a function of satellite stellar mass, but also by visual magnitude assuming a mass-to-light ratio of unity. Given the D17 predictions (gray line and shaded region), M33 could host up to 15 satellites with stellar masses $> 10^{3} \, M_{\sun}$ within the expected errors. All solid colored lines correspond to HSC and all dot dashed colored lines correspond to MC. 

The number of satellites expected to reside roughly within the 50 kpc area observed around M33 by PAndAS is provided by the rows indicating $\rm N_{fields}=23$ in Table \ref{table:HSC} and $\rm N_{fields}=41$ in Table \ref{table:Mega}. These limits are also indicated by the intersection of the orange curves and dotted vertical orange line in Fig. \ref{fig:794kpc}. At the stellar mass of And XXII, M33 is expected to host about three satellites within this area on average. Therefore the discovery of only one potential satellite companion of M33 by PAndAS is lower than the number of luminous satellites calculated to exist within 50 kpc of M33 using the adopted methodology. 

PAndAS surveyed about one-third of M33's virial radius (see Fig. \ref{fig:coverage}). Naively, one would expect that roughly (50 kpc)$^2$/(161 kpc)$^2$ (or 9.6\%) of the satellites predicted in M33's full virial volume to be detected within this area on the sky. However, the number of satellites expected using $K_{\rm los}(50/161)$ is $\sim$53\% of the total expected satellites. Similarly, for a survey radius of 100 kpc, one would naively expect (100 kpc)$^2$/(161 kpc)$^2$ (or 38.6\%) of the total satellites to be detected, but $\sim$92\% of the total expected satellites are predicted to be observed in that survey area. 

The discrepancy between naive expectations and our results is a combination of effects associated with the geometry of proposed observations and the radial distribution of subhalos. Geometrically, a 50 kpc survey radius corresponds to a 50 kpc cylinder cut out of a sphere with a radius of $\rm  R_{vir}=161$ kpc (also known as a spherical ring), which is about 14\% of the sphere's volume compared to the 9.6\% expectation when two cylinders of different radii are considered. Furthermore, the radial distribution of subhalos is not uniform. Generally subhalos follow the radial dark matter density profile of their primary host halo in $\Lambda$CDM. The radial distribution of satellites labeled as luminous owing to reionization tend to be even more centrally concentrated. In other words, the density of satellites per unit volume is highest near the center of a host halo, resulting in more predicted satellites within smaller survey radii than expected for a uniform distribution. 

\subsection{An Extended Survey of M33's Virial Volume}
The PAndAS survey footprint is approximately 35\% of M33's virial volume (if $\rm D_{M33}=794$ kpc) when the full footprint is accounted for, so while we calculate that 154 fields and 254 fields with HSC and MC, respectively, are required to survey the full projected area of M33's virial volume, only 100 fields with HSC and 165 fields with MC would be necessary to fill in this region. We acknowledge that this number of fields is not trivial to acquire, observe, and process. If one were to survey a radius of only 100 kpc around M33 rather than extending to its virial radius, approximately 50 (HSC) and 100 (MC) new fields of data would be necessary since the PAndAS area is about 50\% of this region. About 90\% of the total M33 satellite population is expected to reside within 100 kpc.

\begin{figure*}
\includegraphics[scale=0.6, trim=15mm 0mm 0mm 0mm]{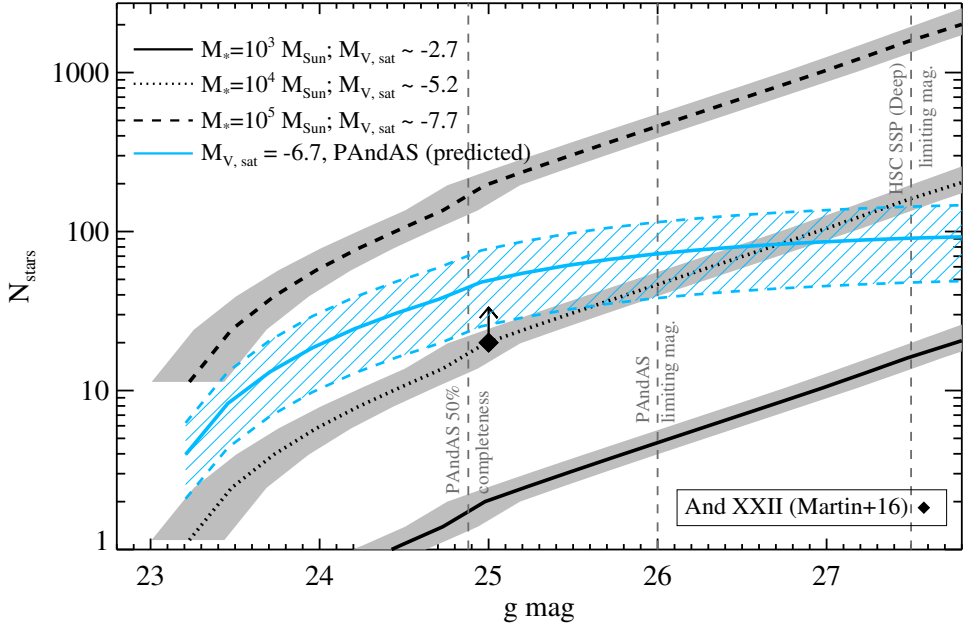}
\caption{The predicted number of resolved stars per M33 satellite as a function of g-band magnitude. The black lines and gray shaded regions show the number of resolved stars for satellites of different masses or $\rm M_V$ across the range of M33 distances considered in this work. The blue line and hashed region indicates the number of resolved stars expected for a satellite of $\rm M_V=-6.7$, the measured magnitude of And XXII (M16), corrected for the PAndAS completeness limits. These calculations assume a [Fe/H]=-2.0 stellar population with an age of 12 Gyr and an IMF power-law slope of -2.0. The black diamond shows the actual number of red giant branch stars observed in the PAndAS survey (M16). The observed number of And XXII stars is approximately consistent with the luminosity function predictions. }
\label{fig:starcounts}
\end{figure*}

Fig. \ref{fig:starcounts} shows the number of resolved stars in a single M33 satellite at various depths in the g-band. These calculations are for an [Fe/H] = -2.0 stellar population with an age of 12 Gyr and a power law initial mass function with a slope of -2.0. Predictions for satellites of various stellar masses (and $M_V$) at the distance of M33 are shown in the black lines and gray shaded regions, where the shaded regions span the various M33 distances discussed in this work. The blue lines and  hashed area show how many stars are expected at various g-band magnitude limits for an $M_V=-6.7$ satellite corrected for the PAndAS completeness limits (see M16). 

The black diamond indicates the resolved red giant branch stars in And XXII from M16. Our predictions for And XXII are in agreement with this simple luminosity function analysis. The vertical dashed gray lines represent the PAndAS 50\% completeness limit, PAndAS limiting g-band magnitude and the HSC Subaru Strategic Program (SSP)\footnote{See \url{http://hsc.mtk.nao.ac.jp/ssp/survey/}.} limiting g-band magnitude, respectively. When we change the default age of 12 Gyr and [Fe/H]=-2.0 to luminosity functions with the same age and [Fe/H]=-1.0, or with the same metallicity and an age of 5 Gyr, the predicted number of stars changes by 20 per cent at most.

With HSC on the 8.2m Subaru Telescope, reaching the PAndAS magnitude limit of $g\sim26$ requires only $\sim4$-minute exposures, and $i\sim25.5$ can be achieved in $\sim12.5$-min exposure time. Based on these exposure times, we estimate that a survey with Subaru/HSC covering 100 fields to PAndAS depth would require only $\sim4$ nights of observing time. As illustrated in Figure~\ref{fig:starcounts}, observations extending to the depth of the HSC-SSP Deep fields ($g\sim27.5$) could enable detection of $\sim20$ stars in satellites as faint as $M_* = 10^3~M_{\sun}$. HSC-SSP has required exposures of 1.4/2.1 hours per field in $g/i$ to reach these depths. This is technically feasible, and could increase the expected number of M33 satellites by a factor of $\sim3$ compared to the PAndAS depth. However, more reasonable exposure times of $\sim25$ and 55 minutes would reach depths of $g\sim27$ and $i\sim26.3$, enabling detection of most satellites to a few times $10^3 \, M_{\sun}$. Such a survey would likely require $\sim12-14$ nights to cover 100 HSC fields. The fainter satellites would make the model comparisons more statistically robust, and in addition these faintest galaxies could be the relics of the reionization era and the first populations of stars. 

The imagers discussed thus far are both on ground-based telescopes, but upcoming space observatories will be far better equipped to tackle the region surrounding M33. WFIRST is expected to have a wide field imager with a field of view nearly 100 times that of the Hubble Space Telescope's IR channel. With a 0.28 square degree field of view, a region of 125 square deg (or a 160 projected kpc radius around M33) would require 448 fields. The 2.4m aperture of WFIRST will be able to reach the same depths as HST in similar exposure times. For example, to reach $M_V=-7$ (or $\sim$17.5 in the V-band), short exposures of only 36 seconds could reach 6 magnitudes deeper than this in a total of $\sim$4 hours. Of course, this does not account for the additional overheard and slew time. Regardless, this would still be a small observing program for an observatory with the capabilities of WFIRST. A survey around M33 may also be suitable for proposed missions such as the Habitable Exoplanet Imaging Mission (HabEx), which will be able to observe resolved stellar populations in nearby galaxies.

The predictions listed in Tables \ref{table:HSC} and \ref{table:Mega} only consider the satellites that are bound to M33. The total number of galaxies observed around M33 would actually be a combination of M33 satellites and M31 satellites that happen to be near M33 since M33's halo is encompassed by M31's halo. These background M31 satellites are not included in our predictions but may be observed in extended surveys around M33. Adopting a mass of $1.5-2 \times 10^{12} \, M_{\sun}$ for M31, at a distance of about 200 kpc away, approximately a dozen M31 satellites with stellar mass $\geq 10^3 \, M_{\sun}$ may coincide with the virial area of M33. If new satellites are detected around M33, additional kinematic information will be necessary to determine which host galaxy the satellites are bound to, similar to the case of And XXII. 

Our results also assume that M33 has evolved in isolation and that its potential satellites have not endured strong tides from the larger environment surrounding M33. However, the past orbital history of M33 is not well-constrained, so these results are subject to change if M33 had any recent interactions with other galaxies, such as its host, M31. In the next section, we discuss plausible M33 orbital histories from the literature and the implications of such histories for its predicted satellite population.

\section{The Orbital History of M33}
\label{sec:orbitalhistory}

While the results in Section \ref{subsec:fullsurvey} are applicable to M33 at the time it first fell into M31's halo, the number of surviving M33 satellites is highly sensitive to M33's recent orbital history. In P17A, cosmological analogs of M33-mass subhalos around M31-mass hosts strongly favored more recent infall times with about 70\% of all M33 analogs exhibiting infall times in the last 4 Gyr. Infall time was defined as the first time a subhalo crossed into the virial radius of its host while moving inwards. The trajectory after infall, however, can vary significantly. Constraining the orbital history of M33 is therefore key to understanding and predicting its surviving satellite population.

Computing the orbital history of halo substructures requires knowledge of several basic parameters. To start, the distance to both the host and satellite are crucial initial conditions (for analytic integrations) or as consistency checks (for the output of an N-body simulation). M31's distance is constrained to $785\pm25$ kpc \citep{mcconnachie05} and estimates preceding this do not stray far from this range. M33's distance, however, spans a wider range of values from about 794 to 968 kpc \citep{mcconnachie04, u09} and can alter solutions for the past trajectory of M33.

In addition to distance, the 3D velocity of both the host and satellite are necessary. The proper motion of M33 was measured by tracking the motion of water masers using the \emph{Very Long Baseline Array} (VLBA) \citep{brunthaler05}.  M31's transverse motion has only been measured recently with the Hubble Space Telescope (HST) and \textit{Gaia} \citep{sohn12, vdm12ii, vdm18}. Prior to that, calculations of M33's orbit around M31 explored a wide range of plausible M31 tangential velocities. In Section \ref{subsec: morphorbits}, we discuss two studies that calculate M33's orbit without a measured value for M31's tangential velocity. In Section \ref{subsec:myorbit}, we discuss studies that use directly measured and estimated values for M31's tangential motion to do the same. We also discuss the implications of these orbital histories on M33's satellite population and argue that the number of satellites discovered around M33 can narrow the potential solutions for M33's orbital history.

\subsection{M33's Morphologically Motivated Orbital History}
\label{subsec: morphorbits}
Two recent studies have used the past orbit of M33 to find an explanation for its current morphology. Here we are only referring to the morphology in the outer regions of M33's disks. Knowledge of a warp in the outer HI disk of M33 was first reported in \citet{rogstad76} and later by \citet{corbelli97, putman09, corbelli14, kam17}. While \citet{rogstad76, corbelli97} both concluded that the tidal force from M31 was not strong enough to induce this warp, more recent studies have revisited these claims. 

For example, P09 showed that the gaseous M33 warp extends to nearly 22 kpc from the galaxy's center. As a result, they explore the possibility of a tidal interaction between M33 and M31 to explain such features by integrating orbits backwards in time. In the absence of a measurement for M31's tangential velocity components, they explore a range between -200 km s$^{-1}$ and 200 km s$^{-1}$. For distances to M31 and M33, they adopt 770 kpc and 794 kpc, respectively. An M31 virial mass $> 2.5 \times 10^{12} \, M_{\sun}$ at $z=0$ that decreases exponentially as a function of redshift is used throughout the P09 analysis. M33 is treated as a point mass of $\sim10^{10} \, M_{\sun}$ that evolves in the presence of M31 and the MW, where the MW is also modeled as a point mass of $10^{12} \, M_{\sun}$. P09 concludes there is a 60\% probability that M33 and M31 did reach within 100 kpc of each other about 3 Gyr ago, which yields an M33 tidal radius of 15 kpc. As a result, the tidal and ram pressure forces acting on M33 as it moves through M31 at a close distance are proposed to give rise to the asymmetries in M33's gaseous disk.

A warp in the stellar disk of M33 was discovered by the PAndAS survey (M09). The stellar warp has a similar orientation to the gaseous warp and extends about 30 kpc from its center. In an effort to reproduce the stellar morphological features observed in the PAndAS survey, M09 created a suite of N-body simulations of the M33-M31 system to search for an orbital history that leads to a tidal interaction. These simulations aim to reproduce the stellar morphologies of M33 and M31, so they do not follow the evolution of neutral gas in either galaxy.

Using over 6 million particles, M33 and M31 are each modeled with a halo, disk and bulge component. The simulations use total masses of $2.56 \times 10^{12} \, M_{\sun}$ and $8.27 \times 10^{10} \, M_{\sun}$ for M31 and M33 respectively. The tangential velocity of M31 had yet to be measured, so a range of plausible velocities were explored to match the present day position and velocity of M33 in addition to the morphological features \citep[see also][]{chapman13}. M09 finds several representative orbits of an M33-M31 encounter, one of which suggests that a pericentric passage at 55 kpc about 3 Gyr ago achieves the desired kinematics and observed features. Both the gaseous and stellar M33 warps can be accounted for by M09 and P09's proposed orbits, which together suggest M33 completed a pericentric passage at < 100 kpc in the last 3 Gyr. 

Interestingly, a pericentric approach as close as 50 kpc could strongly truncate or perhaps destroy the gaseous disk of M33 \citep[see][for a study of the LMC's disk as it moves through the MW's circumgalactic medium]{salem15}, but this is not evident in HI observations of M33. \citet{dobbs18} have also shown that M33's flocculent spiral pattern and velocity field as seen in HI are reproducible in simulations through gravitational instabilities in the stars and gas alone, further supporting that an interaction is not required.

\subsubsection{Implications for M33's Predicted Satellite Population}
\label{subsubsec:implications1}

If M33 has already completed a pericentric passage about M31 in the last 3 Gyr, M31 could have tidally stripped satellites from M33's halo. These stripped satellites may appear to be satellites of M31 today. The strength of tidal forces depends directly on the distance between the two galaxies when M33 is at pericenter. In this section we explore how many satellites would survive a pericentric passage at a range of pericenter distances. Note that these calculations assume that all satellites outside of the calculated tidal radius are completely stripped from M33 and more detailed simulations of these scenarios should be explored.

The tidal radius is computed following Eq. 3 of \citet{vdbosch18a}, which models both the host and satellite galaxy as extended objects:

\begin{equation}
\rm r_{t} = r_{peri}\left[\frac{m(r_t)/M_{host}(r_{peri})}{2 - \frac{d lnM}{d ln R}\rvert _R}\right]^{1/3}.
\end{equation} 

M33 is approximated as a Plummer sphere \citep{plummer11} with a total mass of $2.5 \times 10^{11}\; M_{\sun}$ and a scale length of 20 kpc. M31 has been approximated as an NFW halo \citep{nfw96} and we consider two different masses: $1.5\times 10^{12}\; M_{\sun}$ with concentration $c_{\rm vir}  \equiv r_{\rm vir}/ r_s=9.56$ and $2\times 10^{12}\; M_{\sun}$ with $c_{\rm vir} = 9.36$. For pericentric distances of 50 kpc, 100 kpc, and 150 kpc, we have listed the corresponding tidal radii in  Table \ref{table:tidal} along with M31's mass enclosed at that distance and the fraction of the volume within the tidal radius relative to the virial radius.

\begin{table}
	\centering
	\caption{The tidal radius of M33 at three different pericentric distances: 50 kpc, 100 kpc, and 150 kpc from M31. All tidal radii are calculated assuming a virial mass of $2.5 \times 10^{11}\; M_{\sun}$  and scale length of 20 kpc for M33 where M33 is represented as a Plummer sphere. M31 is modeled as an NFW halo with masses of $1.5 \times 10^{12} \, M_{\sun} \, (2 \times 10^{12} \, M_{\sun}$) and concentrations of 9.56 (9.36). The final row gives the fraction of the tidal radius volume relative to the virial volume of M33, suggesting that if satellites outside of $r_{\rm tidal}$ are stripped, very few are expected to remain bound to M33. } 
	\begin{tabular}{cccc} 
		\hline
			$\rm r_{peri} $ & $\rm M_{M31}(r_{peri})\;$ & $\rm r_{tidal}$ & $\rm \frac{V(r_{tidal})}{V(r_{vir})}$ \\ 
					\, [kpc] & \,  [$10^{11}\, M_{\sun}$] & \, [kpc] & \, \\ \hline
			50 & 3.6 (4.2) & 29.2 (26.9) & 0.05  \\
		        100 & 7.0 (8.6) & 52.6 (48.8)  & 0.16 \\
		        	150 & 9.7 (12) & 73.1 (67.8) & 0.30\\ \hline
	\end{tabular}
	\label{table:tidal}
\end{table}

For a pericentric approach at 50 kpc, only satellites in the inner 5\% of M33's virial volume are expected to remain bound after such an encounter\footnote{In this section, ``bound" refers to whether a satellite can escape the influence of M33's gravitational potential due to M31's tidal forces.}. This yields a tidal radius of 27-29 kpc depending on the mass of M31. From the results in Tables \ref{table:HSC} and \ref{table:Mega} only one to two satellites on average are expected to be in a survey radius of $\sim$30 kpc that reaches the same photometric depth as PAndAS ($M_V \sim -6.5$; $M_* = 10^4-10^5 \, M_{\sun}$). PAndAS surveyed a larger area but only one potential M33 companion was identified. 

At a pericentric passage of 100 kpc, only satellites within 49-53 kpc (16\% of the virial volume) of M33 are estimated to be bound today. This yields two to four satellites with $M_*$ between $10^4 \, M_{\sun}$ and $10^5 \, M_{\sun}$ in a circular area with a radius of $\sim$50 kpc. For a wider pericentric approach of 150 kpc, a tidal radius of 68-73 kpc yields an average of three to six satellites at these detection limits. Fainter M33 satellites ($M_* \sim 10^3 \, M_{\sun}$) are also predicted to exist within these proposed survey radii, but deeper observations would be necessary to robustly detect such faint objects. 

If a total of approximately four or more satellites ($M_* \gtrsim 10^4 \, M_{\sun}$) are discovered and furthermore confirmed as true M33 satellites (i.e. through proper motion analysis), a recent, close (< 100 kpc) tidal encounter \citep[M09, P09,][]{semczuk18} is unlikely under the assumption that the D17 predictions are correct since only one or two satellites are estimated to survive such a close interaction. 

\subsection{M33's Orbital History Using its Current Space Motion}
\label{subsec:myorbit}

\subsubsection{Orbits Using Direct M31 Proper Motion Measurement}
Section \ref{subsec: morphorbits} summarizes M33 orbital histories that aimed to reconstruct the observed morphological structure of the M31-M33 system. P17A computed the orbital history for M33 around M31 using only the current space motion of both galaxies. Such calculations still require assumptions for the mass of both galaxies, however, the focus is shifted to the most statistically common orbital histories for M33. These calculations are made possible by direct measurement of M31's proper motion using the HST \citep{sohn12,vdm12ii}. \citet{sohn12} presents the direct measurement from three fields of data. \citet{vdm12ii} corrects the HST measurement for the internal motion of M31, viewing perspective, and weights this measurement with several indirect methods discussed in \citet{vdMG08}. Both HST derived measurements yield similar M33 orbits (see P17A for more details).

P17A follows a similar methodology to P09 to compute M33's orbit. A three component analytic potential is adopted for M31 but M33 is approximated by a Plummer sphere rather than a point mass. The latter is necessary to approximate the effects of dynamical friction accurately. Rather than choosing just one mass for either galaxy, a range of masses is explored resulting in six different M31-M33 mass combinations. M33's mass range is set by the dynamical mass estimate on the low end and extends up to masses predicted by AM relations. This results in a range from $5-25 \times 10^{10} \, M_{\sun}$ \citep{corbellisalucci,guo11}. For M31, virial masses of $1.5-2 \times 10^{12} \, M_{\sun}$ are considered, similar to our tidal radius calculations. Note that these M31 masses are lower than those adopted in both M09 and P09.

Using the 6D phase space information derived from proper motions of both galaxies, 10,000 orbits are calculated for each mass combination. The 10,000 orbits encompass measurements errors in the distances to M33 and M31, their proper motions, and the measurement errors in the solar quantities, which are used to correct for a galactocentric reference frame \citep{vdm12ii}. Less than 1\% of all orbits achieve a pericentric passage within 100 kpc of M31 during the last 3 Gyr, regardless of M33 mass. If the mass of M31 is increased to $> 2\times10^{12} \, M_{\sun}$, this increases to a few percent of orbits. Instead, most orbital solutions favor a scenario where M33 is on its first infall into M31's halo \citep[like the Magellanic Clouds in the Milky Way;][]{b07}, or it is on a long period orbit where its last pericentric passage was about 6 Gyr ago at a distance of $\sim100$ kpc from M31. These solutions are in agreement with other dynamical studies of LG galaxies \citep{shaya13, vdm12iii}. 

All calculations assume a distance of $\sim 800$ kpc to M33, which corresponds to a separation of about 200 kpc between M33 and M31. Higher M33 distance \citep{u09} measurements would suggest larger separations between M33 and M31 (in excess of 220 kpc). Larger separations provide further support for a first infall scenario. In the event of a long period orbit, higher M33 distances suggest that the pericenter occurred closer to 5.5 Gyr ago or that their separation at pericenter was $>100$ kpc. The orbits of cosmological M33 analogs in P17A that completed a pericentric passage about their host ($\sim$77\% of the sample) exhibited average infall times $t_{\rm inf}= 3.9\pm2.1$ Gyr and a wide range of average distances at pericenter where $r_{\rm peri}=89.8\pm60.2$ kpc. 

\subsubsection{Orbits Using M31 Proper Motion Estimates}
The P17A models use the M31 proper motion measurement from \citet{vdm12ii}, but other values have also been reported through indirect methods. \citet[][hereafter S16]{salomon16} inferred M31's proper motion using the one-dimensional kinematics of its satellites. This yields an M33 orbital history in which it makes a pericentric approach in the last 2-3 Gyr, but only at distances of 140-175 kpc on average (P17A). At these separations, it is not likely that M31 can induce any strong tidal features such as the warps seen in M33's disks.

More recently, \citet{semczuk18} has explored the range of relative velocity vectors derived from the S16 M31 proper motion estimates. Similar to our P17A analysis, \citet{semczuk18} conclude that the \citet{vdm12ii} M31 proper motion measurement does not favor a recent tidal interaction between M33 and M31. They instead focus on the S16 results, which are more likely to result in a recent encounter between the galaxies within the estimated error space. By varying the initial magnitude and direction of the S16 velocity vector, \citet{semczuk18} recovers orbits with pericentric distances < 100 kpc in the last 2 Gyr, settling on a fiducial orbit with a pericentric distance of 37 kpc at 2.7 Gyr from the start of their simulation. Note that this distance at pericenter is even smaller than that suggested by both P09 and M09. It is also closer than the predicted pericentric distance for the LMC relative to the MW \citep{b07}. 

Using this fiducial orbit, \citet{semczuk18} run an additional N-body/SPH simulation of the M33-M31 system and reproduce features similar to the gaseous warps and stellar streams in M33 only when an M31 hot halo is included. While \citet{semczuk18} perform a thorough analysis of the S16 M31 proper motion error space, the S16 errors are about twice as large as those in \citet{vdm12ii}. The relative velocity vector corresponding to the best match projection in neutral hydrogen is 1-2$\sigma$ in each component from the average S16 results and up to 4-5$\sigma$ from the \citet{vdm12ii} values. Such discrepancies in the orbital history of M33 clearly demonstrate the need for additional direct measurements of M31's proper motion. Third epoch measurements with HST may be able to provide a factor of two to three improvement for M31's tangential velocity. 

Independent proper motions for M33 and M31 have already been measured with \textit{Gaia} DR2 \citep{gaiadr2a, vdm18}. These results agree with the \citet{vdm12ii} M31 results at $< 1\sigma$. For M33, the Gaia DR2 proper motion is in agreement with the VLBA \citet{brunthaler05} proper motion at the $1\sigma$ level. The Gaia DR2 proper motions are an independent consistency check that the previously measured M33 and M31 proper motions are accurate. When the M31 DR2+HST weighted average and the M33 DR2+VLBA weighted average are used to compute the orbital history of M33 similar to P17A, a first infall scenario is unanimously preferred. 

\subsubsection{Implications for M33's Predicted Satellite Population}
In Section \ref{subsubsec:implications1}, we conclude that the discovery of four or more confirmed satellites could suggest a close tidal interaction between M33 and M31 \citep[M09, P09,][]{semczuk18} is unlikely, leaving only a long-period orbit or first infall scenario (P17A). 

If M33 made its closest approach to M31 around 6 Gyr ago at a distance of 100 kpc, some tidal stripping would be expected and this could reduce the number of bound M33 satellites observed today (see Section \ref{subsec: morphorbits}). Only two to four surviving satellites are predicted to remain bound to M33 in such circumstances. Recall that the interaction history between M33 and M31 is typically used to explain M33's morphology. If M33 experienced a pericentric passage at larger distances (> 100 kpc) as suggested in P17A, tidal interactions alone are unlikely to be the origin of M33's warps. Interactions between satellites of M33 and M33 must therefore be invoked. Our calculations confirm that M33 would retain a fraction of its satellite population in these orbital scenarios and subsequently that tidal forces owing to M31 plus interactions with satellites could lead to its current morphological asymmetries. These conclusions do not account for the potential relaxation of the gaseous and stellar disk after this type of tidal interaction, thus simulations should be carried out to determine how long-lived such features are on average relative to a 6 Gyr orbital period.

If M33 is on first infall and has evolved in isolation for a majority of its lifetime, its satellite population is expected to be almost fully intact. A full virial volume survey could result in four to eight satellite galaxies with $M_* = 10^4-10^5 \, M_{\sun}$ if the D17 predictions hold. Up to seven additional, fainter satellites ($M_* < 10^4 \, M_{\sun}$) may also be detected with a survey reaching greater photometric depth than PAndAS. Due to the higher concentration of satellites in the inner region of M33's halo, even a 100 kpc survey radius (twice that of the PAndAS survey) is expected to yield $\gtrsim 90 \%$ of the predicted M33 satellite population. 

Greater than four M33 satellite candidates would provide further evidence for a first infall scenario. On the other hand, a deficiency of satellites may confirm a recent, pericentric approach of M33 around M31. Alternatively, it could suggest that the D17B methodology needs revision. Recall that the determination of luminous satellites from dark subhalos is sensitive to the $M_* - M_{\rm halo}$ relationship and the effects of suppressed star formation due to reionization. Different AM models or a modification of the reionization implementation may alter the resulting satellite population. Finally, the $\Lambda$CDM cosmology itself could be flawed and perhaps other types of dark matter (i.e. warm, hot, self-interacting) may need to be considered.

\section{Discussion}
\label{sec:discussion}
The predictions provided in this work represent the number of satellites expected around an LMC or M33-mass host at $z=0$ after accounting for the accretion history and the potential group pre-processing that may have taken place prior to infall \citep[see][]{wetzel15a}. M33 could have up to 15 satellites down to a limiting magnitude of $\rm M_{V}= -3$ today according to the GK17 AM model in D17B. A more recent study suggests that LMC-mass hosts could have $\sim$25 surviving satellites down to $\rm M_V=-3$ if reionization occurred at $z\sim6$ \citep{bose18}, indicating that these predictions strongly depend on the choice of AM method and the time of reionization. 

Simulations are also known for the over-disruption of subhalos owing to numerical effects associated with simulation softening length choices in cosmological simulations \citep{vdbosch18b} and tidal effects due to the presence of a galactic disk \citep{gk17b}. AM techniques are calibrated against simulations to match with observations and incorporate such inaccuracies in the $M_* - M_{\rm halo}$ relationship itself. These processes are not expected to underestimate the satellite population for a single AM technique, but may add additional variation to this type of analysis across many AM methods. 

Additionally, the cosmological zoom simulations used in the D17 formalism and \citet{bose18} do not include baryonic physics, but the co-evolution of baryons and dark matter is known to impact the abundance and properties of dark matter substructures \citep[e.g.][]{gk17b, dolag09, brooks14, wetzel16, sawala17}. For example, \citet{gk17b} and \citet{sawala17} have recently studied the impact of baryons in cosmological zoom simulations of MW-like halos using different prescriptions for star formation and feedback and both studies conclude that the dark matter only simulations overpredict the number of subhalos within 300 kpc of the simulated MW-like halos by factors of 1.2-1.5 relative to the full hydrodynamical simulation counterparts. This difference is largely attributed to the gravitational influence of the host galaxy's disk. Within 50 kpc of the host, subhalos counts are overpredicted by a factor of 1.75-2 \citep{sawala17} or 3-4 \citep{gk17b}.

If the same disparity holds for halo masses an order of magnitude less massive ($\sim10^{11} \, M_{\sun}$), the subhalo populations for galaxies like M33 would also be overpredicted by similar amounts. The subhalo counts themselves determine the SHMF and are important for establishing abundance matching relations between observations and simulations, but the time of reionization and how it is implemented also plays a key role in determining which of the subhalos host luminous satellites. As a result, an overabundance of subhalos in dark matter only simulations does not directly correlate to an overprediction of luminous halos. While the combination of these effects may affect the number of predicted satellites around galaxies like M33, it is unclear by exactly how much once both numerical effects and contributions from baryonic physics are reconciled and this should be studied in further detail.

Regardless of these phenomena, if M33 is just now reaching the closest distance to M31 ever, it is likely to have been host to more satellites prior to infall than these studies suggest. Below, we discuss the satellites of satellites hierarchy, its implications for the merger history and morphological evolution of an M33-mass galaxy, and the mass function of the predicted M33 satellite population.

\subsection{Implications for Satellites of Satellites}
If the M33 orbital history presented in Section \ref{subsec:myorbit} is correct and M33 did not have a recent, close tidal interaction with M31, alternate explanations for M33's gas and stellar disk warps are required. One potential solution is that satellites themselves could induce such warping through close interactions, high speed flybys, collisions, or mergers. 
 
Simulations of discy dwarf galaxy hosts with $\rm M_{vir} = 5.6\times 10^{10} \, M_{\sun}$ and dark subhalos with 5-20\% of the host mass have illustrated that subhalos can alter the kinematics and structure of the dwarf hosts during interactions \citep{starkenburg16b}. They find that dark satellites on radial orbits can especially cause structural changes in the host galaxy that manifest as asymmetries in both the gas and stars. For slightly higher mass host halos like M33, similar processes between dark matter dominated dwarfs or dark satellites may also lead to morphological features that mimic those produced by tidal interactions. The mass ratios of such encounters, however, would have to be low (i.e. halo mass ratios $\lesssim 1/100$) and at distances that would not perturb the innermost regions of M33's gaseous and stellar disks. 

For example, \citet{semczuk18} provides a basic analysis of whether And XXII could provide the tidal impact necessary to induce M33's warps. They conclude that And XXII would have to reach very close distances ($\sim$5 kpc) to induce such features at which point M33's disk may be affected. We emphasize that while one close interaction between M33 and a satellite may not be enough to induce the warps and debris in and around M33, several interactions with one or more satellites could amount to its current morphology.
 
Following $\Lambda$CDM hierarchical assembly, M33 should have experienced several mergers already. The cumulative merger histories of $\sim$100 massive satellite galaxy analogs ($\rm M_{halo} > 10^{11} \, M_{\sun}$, P17A) residing in M31-mass halos ($0.7-3 \times 10^{12} \, M_{\sun}$) in the lllustris-1 cosmological simulation \citep{vogelsberger14a, nelson15} show that the median number of galaxy-galaxy mergers throughout a massive satellite galaxy's lifetime is $3\pm1.5$ \citep{rg15}. This excludes any mergers between galaxies that cannot be traced back to two different friends-of-friends groups, but it does include all mergers with stellar mass ratios > 1/10. A floor of ten stellar particles ($M_* \sim 10^7 \,M_{\sun}$) for the smallest progenitor in each merger is imposed, thus these estimates will not include merger events involving ultra-faint dwarf galaxies, for example.

Recent work on dwarf-dwarf mergers has shown that merger remnants are biased towards larger distances in host galaxy halos since mergers typically occur before infall thereby resulting in more recent accretion times \citep{deason14b}. If M33 is indeed on its first infall into M31's halo with an accretion time of 2-4 Gyr ago as suggested in P17A, this picture is consistent with M33's current morphology and separation of about 200 kpc from M31, or beyond half of the virial radius of M31. 

\citet{jethwa16} have suggested that a total of $70^{+30}_{-40}$ satellites between $-7 < M_V < -1$ could have evolved within the virial radii of either of the Magellanic Clouds prior to infall into the MW's halo \citep[see also][]{dwagner15}. If M33 has a similar or greater halo mass than the LMC, it has likely been host to tens of satellites throughout its lifetime as well (once the contribution of satellites associated with an SMC-mass companion are subtracted). The number of mergers predicted using Illustris-1 plus the number of surviving satellites expected to reside around an LMC-mass galaxy today \citep[D17B,][]{bose18} are roughly consistent with the lower limit presented in \citet{jethwa16}. 

\subsection{The Lack of Bright M33 Companions}
Much of our analysis focuses on satellites in the ultra-faint dwarf galaxy regime. However, $\Lambda$CDM suggests that even host galaxies with stellar masses comparable to the LMC or M33 should have fairly smooth satellite mass functions such that they host at least one $M_* \sim 10^6 \,M_{\sun}$ satellite, resulting in a three orders of magnitude difference between host and satellite stellar masses at most. This gap is often referred to as the stellar mass gap statistic \citep[see][]{deason13b}.

An M33 satellite of this mass and brightness ($M_* \gtrsim 10^6 \,M_{\sun}$ and $-9 > M_V > -10$) would have been detected in surveys such as SDSS, yet none have been found. If we assume that And XXII or another satellite of roughly the same stellar mass is M33's next most massive satellite, this leads to a four or five orders of magnitude difference between the stellar mass of M33 and its brightest satellite, suggesting that this is possibly a small-scale $\Lambda$CDM problem. 

Recently discovered dwarfs around the Large Magellanic Cloud also lead to a fairly substantial stellar mass gap for the LMC. There, the difference in the LMC's stellar mass and that of its next most massive satellite is almost six orders of magnitude (see D17B and references within). Unfortunately, the complex orbital history of the LMC and SMC \citep[i.e. their binary nature and recent infall into the MW's halo; see][]{b07,besla12,k13, diaz11} makes it difficult to determine whether this stellar mass gap is long-lived or if interactions between the Magellanic Clouds have changed their total satellite populations and therefore the stellar mass gap characteristic over time.

First results from the MADCASH survey have yielded one satellite galaxy around NGC 2403, a dwarf spiral at 3.2 Mpc with about twice the LMC's stellar mass \citep{carlin16}. The dwarf companion (MADCASH J074238+652501-dw) is estimated to have a stellar mass of $\sim10^5 \, M_{\sun}$ and a previously known satellite, DDO 44 has a stellar mass of approximately $6\times 10^7 \, M_{\sun}$ \citep{karachentsev13}. The presence of DDO 44 leads to only a two orders of magnitude stellar mass gap for this system, which is more consistent with $\Lambda$CDM expectations. However, Besla et al. 2018 (submitted) find that less than 1\% of isolated dwarf pairs in both SDSS and cosmological mock catalogs are stellar mass analogs of the Magellanic Clouds, suggesting that mass gaps of about one order of magnitude are very uncommon. Further results from the MADCASH survey will increase the sample of LMC stellar mass hosts and their satellite populations, helping to decipher whether stellar mass gaps are truly in contention with $\Lambda$CDM. A larger sample may also inform our knowledge of how host environment may affect the properties and number of satellite companions around LMC/M33 stellar mass galaxies.

\section{Conclusions}
\label{sec:conclusions}
We have tabulated the number of satellites expected to reside within the virial radius of M33 following the D17 framework to motivate a concentrated search for M33 satellites beyond the existing PAndAS survey region. To date, there is only one potential M33 candidate satellite, And XXII, but it is unclear based on its one-dimensional kinematics whether And XXII is bound to M31 or M33. A proper motion measurement for And XXII would make it possible to derive a full orbital history in the combined M31+M33 gravitational potential and possibly determine if it is a true satellite of M33 or just another member of the larger M31 system. 

The discovery of additional M33 satellites or the lack thereof could further constrain the orbital history of M33. The common orbital histories in the literature have different implications for the M33 satellite system as one \citep[M09, P09,][]{semczuk18} suggests it had a recent tidal interaction with M31 that occurred at close enough separations to induce warps in M33's stellar and gaseous disk and potentially strip satellites away from M33's halo. On the other hand, the current space motions of M33 and M31 favor a scenario where M33 was only recently accreted into M31's halo and is moving towards its first pericenter or a scenario in which M33 already completed its first pericentric passage around M31 at about 6 Gyr ago and at separations $\gtrsim$100 kpc \citep[P17A,][]{vdm18}. If M33 is on its first infall, its satellites are expected to remain bound and the satellites themselves may explain M33's morphology. A wide pericentric passage may strip away some outer M33 satellites but a majority of the satellites are still expected to remain dynamically stable.

While a survey encompassing the full virial area of M33 on the sky that reaches the same depth as the PAndAS survey would provide the most complete picture of the M33 satellite population, a survey of this magnitude would not be trivial. A survey extending to 100 kpc in projection from the center of M33 (or twice the PAndAS survey radius), however, would yield up to eleven M33 satellites on average with $M_* \geq 10^3 \, M_{\sun}$ (or about five satellites at the PAndAS photometric limits) if the D17 predictions are correct. This is approximately $\sim$90\% of the total predicted M33 satellite population. The PAndAS survey area already observed about 50\% of this region, so a ground-based imager like HSC with a 1.5$^{\circ}$ field of view diameter would need to observe about 50 fields of new data. In the era of WFIRST, this proposed survey would be a small program requiring about 224 fields and short exposure times of about 36 seconds each.

We conclude that the discovery of four or more new M33 satellites would strongly disfavor the recent, close tidal interaction scenario. Greater numbers of satellites would provide further evidence of a first infall scenario or a long-period orbit at larger pericentric distances. Upwards of about six M33 candidate satellites would permit only a first infall scenario for M33 given the measured positions and velocities of M33 and M31.

In addition to dwarf satellite galaxies, extending the PAndAS survey region may also result in the discovery of additional globular clusters. Studies of the inner \citep[< 10 kpc;][]{sanroman10} and outer \citep[10-50 kpc;][]{cockcroft11} regions of M33 concluded M33 has a much lower globular cluster surface density than M31, especially in its outermost regions. These results suggest that some M33 globular clusters may have been accreted by M31 through recent tidal interactions or that M33's accretion history is calmer than M31's, which could mean it evolved in a more isolated, low density environment. The latter would also support a first infall scenario for M33. 

The disparity between the orbital histories presented in M09, P09, and \citet{semczuk18} compared to P17A and \citet{vdm18} leads to further discussions on the history of the larger M31 system. For example, a first infall scenario for M33 suggests that it entered M31's halo around the same time that the progenitor of M31's Giant Southern Steam was moving towards the center of M31 for the first time \citep{ibata04,font06,fardal06}. The implications of two massive satellites entering the halo of an M31-mass galaxy have yet to be explored and will be the topic of future work. 

Additional deep surveys of M33 and LMC stellar mass analogs residing within the halos of intermediate mass galaxies like the MW and M31 will further our understanding of satellite populations around low mass host galaxies in the era of WFIRST and LSST. These data will also allow us to comment further on whether the lack of bright companions around M33 is a true obstacle for $\Lambda$CDM. Since approximately one-third of all MW-mass halos host a massive satellite like the LMC or M33 \citep[P17A,][]{bk11, tollerud11b, robotham12}, the `satellites of satellites' phenomena is an intriguing way to increase our understanding of the faint end of the galaxy luminosity function overall.


\section*{Acknowledgements}
EP is supported by the National Science Foundation through the Graduate Research Fellowship Program funded by Grant Award No. DGE-1746060. We thank Nicolas Martin for providing the PAndAS footprint data, as well as Roeland van der Marel for insights on the capabilities of WFIRST. Many thanks to Gurtina Besla, Beth Willman, Nicolas Garavito-Camargo, and Ragadeepika Pucha for their helpful comments on this project. We are grateful to the Lorentz Center for hosting the {\em Large Surveys of the Great Andromeda Galaxy} workshop which inspired this project.

\section*{Software}
This research utilized: \texttt{IPython} \citep{ipython}, \texttt{numpy} \citep{numpy}, \texttt{scipy} \citep{scipy}, \texttt{astropy} \citep{astropy}, and \texttt{matplotlib} \citep{matplotlib}.



\bibliographystyle{mnras}
\bibliography{masterrefs}

\bsp	
\label{lastpage}
\end{document}